\documentclass[12pt,preprint]{aastex}

\def\arcsecpoint{$''\!.$}
\def\deg{$^{\rm o}$}

\def\gtsim{\raisebox{-.5ex}{$\;\stackrel{>}{\sim}\;$}}

\shortauthors{Crenshaw et al.}
\shorttitle{Mass Outflow from NGC~4151}

\begin{document}

\title{Mass Outflow from the Nucleus of the Seyfert 1 Galaxy
NGC~4151\altaffilmark{1}}

\author{D.M. Crenshaw\altaffilmark{2} \& S.B. Kraemer\altaffilmark{3,4}}

\altaffiltext{1}{Based on observations made with the NASA/ESA Hubble Space
Telescope, obtained from the Data Archive at the Space Telescope Science
Institute, which is operated by the Association of Universities for Research in
Astronomy, Inc., under NASA contract NAS 5-26555.}

\altaffiltext{2}{Department of Physics and Astronomy, Georgia State 
University, Astronomy Offices, One Park Place South SE, Suite 700,
Atlanta, GA 30303; crenshaw@chara.gsu.edu}

\altaffiltext{3}{Institute for Astrophysics and Computational Sciences,
Department of Physics, The Catholic University of America, Washington, DC
20064; kraemer@yancey.gsfc.nasa.gov}

\altaffiltext{4}{Exploration of the Universe Division, Code 667, NASA's
Goddard Space Flight Center, Greenbelt, MD 20771}

\begin{abstract}

We present an analysis of UV and optical spectra of NGC~4151 obtained at high
spectral and angular resolutions with the {\it Hubble Space Telescope's} {(\it
HST's)} Space Telescope Imaging Spectrograph (STIS). We identify a kinematic
component of the emission lines that has a width of 1170 km s$^{-1}$ (FWHM),
intermediate between those from the broad and narrow (emission) line regions
(BLR and NLR). We present evidence that these emission lines arise from the same
gas responsible for most of the high-column UV and X-ray absorption (component
``D+E'') that we see in outflow at a distance of $\sim$0.1 pc from the central
nucleus. The gas in this intermediate-line region (ILR) shields the NLR and has
a global covering factor of $\sim$0.4, based on the observed C~IV fluxes,
indicating mass outflow over a large solid angle centered on the accretion
disk's axis. A large transverse velocity ($v_T \gtsim 2100$ km s$^{-1}$)
compared to the radial velocity centroid ($v_r = -490$ km s$^{-1}$)
indicates that the kinematics is dominated by rotation at this distance, but has
a significant outflow component. The mass outflow rate at 0.1 pc is
$\sim$0.16 M$_{\odot}$ yr$^{-1}$, which is about 10 times the accretion rate.
Based on physical conditions in the gas and dynamical considerations, models
that invoke magnetocentrifugal acceleration (e.g., in an accretion-disk wind)
are favored over those that rely on radiation driving or thermal expansion as
the principal driving mechanism for the mass outflow.

\end{abstract}

~~~~~

\section{Introduction}

Mass outflows of ionized gas from active galactic nuclei (AGN) are detected in
the form of UV and X-ray absorption lines that are intrinsic to the AGN and
blueshifted with respect to the systemic velocities of their host galaxies.
While the radial velocities and physical conditions of the outflows have been
characterized through intensive ground- and space-based observations over the
past decade, their origin(s) and means of outward acceleration are still not
well understood. The most popular dynamical models invoke accretion-disk winds
and one or more acceleration mechanisms, including radiation driving, thermal
expansion, and/or magnetocentrifugal winds (Crenshaw, Kraemer, \& George 2003,
and references therein; Proga 2003; Chelouche \& Netzer 2005; Everett 2005).
Another possible source of the outflows is the putative obscuring torus (Krolik
\& Kriss 2001), although it is possible that the torus is just the outward
extension of a dusty accretion-disk wind (K\"{o}nigl \& Kartje 1994; Elitzur \&
Shlosman 2006).

To distinguish between the dynamical models and determine the nature and origin
of the mass outflows, we need to know the locations and complete kinematics
(transverse as well as radial velocities) of the intrinsic absorbers. Several
in-depth studies of Seyfert galaxies, the nearest bright AGN, have placed the
absorbers at distances of tenths to tens of parsecs ($\sim$100 -- 10,000 light
days) from the central continuum source (i.e., accretion disk plus X-ray corona)
(Crenshaw \& Kraemer 2005, and references therein). On the other hand, the
high-ionization broad (emission) line region (BLR) is much closer, typically at
a distance of only 3 -- 10 light days in these same Seyfert galaxies (Peterson
et al. 2004; Metzroth et al. 2006). Furthermore, the radial velocities of the
absorbers are nearly always $<$ 2000 km s$^{-1}$ (Crenshaw et al. 2003), whereas
the BLR velocities typically extend out to at least 10,000 km s$^{-1}$. Thus, we
have suggested that most of the intrinsic absorbers observed in Seyfert galaxies
are located in the inner narrow (emission) line region (NLR) (Crenshaw \&
Kraemer 2005), based on similarities in locations and velocities, although this
certainly does not rule out an origin much closer to the nucleus.

If we want to detect and characterize mass outflows close to the nucleus, the
best candidate is probably NGC~4151, the nearest ($cz =$ 995 km s$^{-1}$) and
apparently brightest Seyfert 1 galaxy. It's UV spectrum shows intrinsic
absorption lines, spanning a wide range in ionization (O~I to N~V) (Bromage et
al. 1985), from a number of kinematic components that cover a large range in
radial velocity centroid (v$_r =$ 0 to $-$1600 km s$^{-1}$) and velocity width
(FWHM $=$ 15 to 940 km s$^{-1}$) (Kraemer et al. 2001). Component ``D$+$E''
($v_r = -$490 km s$^{-1}$, FWHM $=$ 435 km s$^{-1}$), which is a blend of two
kinematic components identified by Weymann et al. (1997), dominates the
absorption in the UV with its high column density. This component shows variable
ionic column densities that respond to continuum changes and strong metastable
C~III and Fe~II absorption lines (Crenshaw et al. 2000), indicating high number
densities ($n_H \approx$ 10$^{7 - 9}$ cm$^{-3}$) and proximity to the nucleus
(Kraemer et al. 2001).

To further explore the properties of mass outflow in NGC~4151, we obtained
contemporaneous high-resolution UV and X-ray spectra in 2002 May with the {\it
HST} Space Telescope Imaging Spectrograph (STIS), the {\it Far Ultraviolet
Spectroscopic Explorer} ({\it FUSE}), and the {\it Chandra X-ray
Observatory} ({\it CXO}), and retrieved additional archive spectra from
these missions. We give the details of the observations and analysis in
Kraemer et al. (2005, Paper I) and Kraemer et al. (2006, Paper II) for the
X-ray and UV data respectively. To summarize our findings, in order to match
the ionic column densities of D$+$E and the observed covering factors of the
background continuum and line emission in the line of sight ($C_{los}$), we
needed four physical subcomponents (D$+$Ea, b, c, and d) characterized by
different ionization parameters ($U$) and hydrogen column densities ($N_H$),
as shown in Table 1. Based on the ionization parameters and the number
densities ($n_H$) from C III metastable absorption, we found that the D$+$E
absorbers were only $\sim$0.1 pc from the central continuum source. We also
found that D+Ea, the UV absorber with the highest $U$ and $N_H$, was
responsible for the bulk of the X-ray absorption, although we needed to
include an additional, more highly ionized component (``X-High'') to account
for all of the X-ray absorption , which is also listed in Table 1.
Furthermore, we found that the D+E absorbers showed strong evidence for bulk
motion across the BLR. From a {\it CXO} spectrum in 2000 March, we found a
lower $U$ and higher $N_H$ for D+Ea (Table 1) than in 2002 May; the latter
yielded a transverse velocity of $v_T$ $\gtsim$ 2500 km s$^{-1}$ (Paper I).
Between 2001 April and 2002 May, we noticed a change in the $C_{los}$ of D+Ed
(Table 1), which gave $v_T$ $\approx$ 2100 km s$^{-1}$ (Paper II), close to
the previous lower limit.

The proximity of the D+E absorbers to the central SMBH in NGC~4151 and their
large transverse velocities, possibly reflecting orbital motion, suggested to us
that further study of this component could lead to more insight into the
connection between accretion disks and mass outflows in AGN. In particular, the
gas from this component must contribute in some fashion to the emission lines,
and if the contribution is significant and can be isolated, it should
provide additional constraints on the geometry, kinematics, and physical
conditions of the gas. In Paper II, we identified a distinct component of the
He~II and C~IV emission lines in the STIS echelle spectra of NGC~4151 with a
FWHM $=$ 1170 km s$^{-1}$, which is intermediate between that of the C~IV broad
(FWHM $\approx$ 10,000 km s$^{-1}$) and narrow (FWHM $=$ 250 km s$^{-1}$)
components. We noted that the emission from this ``intermediate line region''
(ILR) was absorbed by D+E. Here, we investigate the properties of the ILR in
more detail, using additional emission lines from the six epochs of STIS
high-resolution echelle observation. For convenience, we list the dates of
observations and continuum fluxes at 1350 \AA\ in Table 2.

\section{Analysis}

The available STIS E140M spectra of NGC~4151 provide a unique opportunity to
deconvolve the emission-line components of NGC~4151, due to the high spectral
resolution of the grating (R $=$ $\lambda$/$\Delta\lambda$ $\approx$ 45,000)
and the high angular resolution of {\it HST}. To isolate the ILR contribution,
we selected a low continuum-flux spectrum with the best signal-to-noise, which
was obtained on 2000 June 15. The low state minimizes the contribution of the
BLR to the profile, and the small aperture used for all of the E140M spectra
(0\arcsecpoint2 $\times$ 0\arcsecpoint2) minimizes the contribution from the
NLR. By happenstance, there was a contemporaneous high-resolution observation of
NGC~4151 on 2000 July 2 through a 52 $\times$ 0\arcsecpoint2 slit with the STIS
G430M grating (R $\approx$ 10,000), covering the H$\beta$ and [O~III]
$\lambda\lambda$4959, 5007 lines.

We show the central regions of the emission line profiles from these
observations in Figure 1. The strong blueshifted absorption in the UV resonance
lines (Ly$\alpha$ and the C~IV and N~V doublets) is due to the high-column
absorption component D+E (the other absorption components are identified in
Paper II). H$\beta$ also shows strong D+E absorption in the low continuum flux
state, whereas a STIS G430M spectrum obtained in a high state on 1997 July 17
shows only weak H$\beta$ absorption (Hutchings et al. 2002), consistent with the
weakness or absence of UV low-ionization lines from component D+E in other high
continuum states (Kraemer et al. 2001, Paper II).

Figure 1 shows that the He~II $\lambda$1640 line is unaffected by intrinsic
absorption and has two distinct emission components: one from the NLR (FWHM $=$
250 km s$^{-1}$) and one from the ILR (FWHM $=$ 1170 km s$^{-1}$). There is
likely a contribution from the BLR, but it is too faint to be seen in these data
and is therefore included as part of the continuum fit (spline fits over large
wavelength regions). We fit the profile of He~II with two Gaussians, and used
these profiles as templates to determine the NLR and and ILR contributions to
the other lines. We also detected faint BLR components in the wings of C~IV,
Ly$\alpha$, and H$\beta$, and fit these with splines. To model the observed
emission-line profiles, we reproduced the NLR and ILR templates from He~II at
the expected positions of the lines (including the doublet lines of C~IV
$\lambda\lambda$1548, 1551 and N~V$\lambda\lambda$ 1239, 1243), retaining the
same velocity widths, and scaled them in intensity until we obtained a suitable
match. The fits to the NLR components were straightforward, since they they are
unaffected by absorption, except for that in the cores of the UV resonance lines
due to the host galaxy's disk and/or halo (components F and F$'$ in Weymann et
al. 1997). The fits to the ILR components were also relatively easy to
accomplish, relying on the observed red wings of the profiles for lines with
strong blueshifted absorption. Overall, this method of template scaling provided
excellent fits to the profiles.

As shown in Figure 1, the use of high spatial- and spectral-resolution STIS
spectra at a low continuum state allows us to easily detect and deconvolve the
ILR profiles, which are essentially symmetric around zero km s$^{-1}$.
Inspection of the profiles and fits reveals a couple of interesting results.
1) Component D+E absorbs the emission from not only the continuum and BLR, but
the ILR as well, as can be seen in all of the UV resonance lines. 2) The
maximum blueshifted velocity of D+E ($\sim$ $-$1400 km s$^{-1}$) is close to
the maximum velocity of the ILR, which suggests self-absorption by the ILR.
Hence the ILR contribution resembles a P-Cygni profile, as seen for example in
the H$\beta$ profile.

As discussed in Paper II, if the velocity width of the ILR component is due
primarily to gravitational motion, similar to the case for the BLR (Peterson et
al. 2004), then the FWHM of the ILR profile (1170 km s$^{-1}$) and the mass of
the supermassive black hole (4.1 x 10$^{7}$ M$_{\odot}$, Metzroth et al. 2006)
give a distance from the ILR to the central continuum source of $\sim$0.1 pc.
This is consistent with the distance of the D+E absorbers from the central
source based on their densities and ionization parameters (Paper II).

All of the above suggests a connection between the D+E absorption and the ILR,
including the possibility that the ILR emission lines originate directly from
gas responsible for the D+E absorbers. To test this idea, we measured the
emission-line fluxes from the ILR using the fits in Figure 1. Other UV
emission lines were too weak and/or absorbed to give reliable fluxes. We
determined uncertainties in the fluxes by varying the scale factors applied to
each template until we obtained fits that were clearly no longer acceptable.
Table 3 show the observed line ratios relative to H$\beta$, as well as
reddening-corrected ratios based on the very small reddening of E(B$-$V) $=$
0.02 in our line of sight to the nucleus, which is entirely from our Galaxy
(Crenshaw \& Kraemer 2005). We present photoionization models of the line
ratios in the next section.

The strong C~IV line provides the best opportunity for deconvolving the
emission-line components in the other five epochs of STIS echelle
observations, so that we can characterize the variability in the components.
The other UV lines shown in Figure 1 are noisier and more difficult to
deconvolve in the other epochs, particularly in high states when there is a
strong BLR contribution. An attempt to deconvolve these lines resulted in
error bars that were larger than the measured variations. To determine the
emission-line fluxes of the separate C~IV components in the six epochs of STIS
E140M observations, we followed the same procedure as above; we scaled the
separate C~IV BLR, ILR, and NLR templates from 2000 June 15 until we obtained
matches to the observed profiles.

Figure 2 shows the light curves for the C~IV emission components and the
continuum flux at 1350 \AA. Unsurprisingly, the narrow component remained
constant to within the uncertainties, which is to be expected from emission from
an extended region (constrained by the aperture to be $\sim$14 $\times$ $\sim$14
pc in the plane of the sky and possibly larger in the line of sight). The BLR
shows large amplitude C~IV variations similar to those of the UV continuum
(maximum/minimum flux $\approx$ 10), consistent with previous results from more
intensive UV monitoring that indicate the BLR is small ($\sim$6.8 light days in
diameter, Metzroth, et al. 2006) and therefore responds rapidly to continuum
variations. The ILR shows a positive correlation with continuum flux as well,
although it is not possible to determine a time lag from these undersampled
light curves. However, the timescales of the variations do put an upper limit of
$\sim$140 light days ($\sim$ 0.12 pc) on the radius of the ILR (i.e., from the
time interval between the third and fourth observations), consistent with the
value that we derived above. Interestingly, the amplitude of the ILR light curve
is significantly smaller (maximum/minimum flux $\approx$ 4) than that of the
BLR, which may be be expected from a much larger region in which the amplitude
is reduced by the time-delayed response of different parts of the ILR to
continuum variations (Peterson 1993).

\section{Photoionization Models}

In Papers I and II, we determined that D+Ea has the highest $C_{los}$, $U$, and
$N_H$ of all the UV absorbers. If this subcomponent has a non-negligible global
covering factor, then it must dominate the line emission from the absorbers,
and it is therefore a likely source of the ILR emission. Here, we test
this hypothesis by comparing the predicted emission-line fluxes from our
photoionization models of D+Ea with the observed values from the ILR.

The details of the photoionization models used for this study are given in Paper
I. To summarize, the models were generated using the Beta 5 version of Cloudy
(Ferland et al. 1998), which includes estimated $\Delta$n$=$0 dielectronic
recombination (DR) rates for the M-shell states of Fe and the L-shell states of
the third row elements (Kraemer, Ferland, \& Gabel 2004). We assumed roughly
solar elemental abundances (see Paper II) and that the absorbing gas is free of
cosmic dust, consistent with our previous result that there is essentially no
reddening in our line of sight to the nucleus outside of the Galaxy (Crenshaw \&
Kraemer 2005). As per convention, the models are parameterized in terms of $U$
and $N_{H}$. We modeled the intrinsic spectral energy distribution as a broken
power law of the form $L_{\nu} \propto \nu^{\alpha}$ as follows: $\alpha = -1.0$
for energies $<$ 13.6 eV, $\alpha = -1.3$ over the range 13.6 eV $\leq$ h$\nu$
$<$ 0.5 keV, and $\alpha = -0.5$ above 0.5  keV. We included a low energy
cut-off at $1.24 \times 10^{-3}$ eV (1 mm) and a high energy cutoff at 100 keV.
The luminosity in ionizing photons for the 2002 May epoch was  $Q =
1.1\times10^{53}$ photons s$^{-1}$.

In order to determine whether D+Ea could be the source of the ILR emission, we
scaled the ionization parameter from the 2002 May observation to match the
lower continuum flux observed in 2000 June, which yields $U = 10^{-0.92}$. We
then fixed the other parameters to the 2002 May epoch and ran a new model; we
list the predicted line ratios from this model (``Low-N'') in Table 3.
Overall, the correspondence between the model and observed values is quite
good, considering that no attempt was made to adjust the parameters derived
from the D+Ea absorber to match the ILR ratios. The low [O~III]/H$\beta$ ratio
in particular confirms that we have the correct density ($n_H$).

Table 3 shows that C~IV $\lambda$ 1550 is slightly underpredicted and N~V
$\lambda$ 1240 is too low by a factor of $\sim$2.4. While increasing $n_{H}$ by
a factor of $\sim$ 10 brings the predicted C~IV emission inside the error bars,
the [O~III] $\lambda$ 5007 is then collisionally suppressed. This scenario would
therefore require an additional component of lower density gas to provide the
[O~III] emission. Moreover, N~V is still underpredicted at the higher density.
Although strong N~V emission is often cited as evidence of super-solar nitrogen
abundances (e.g., Hamann et al. 2002), we find that increasing N/H in the models
results in an overprediction of N~IV] $\lambda$ 1486/H$\beta$ (the observed
value is $\sim$1.2 for the BLR, ILR, and NLR combined). The underprediction of
N~V is not unique to this case, as we have encountered it on a number of
occasions when modeling the physical conditions in the NLR (e.g., Kraemer et al.
2000; Kraemer \& Crenshaw 2000).

Having established that it is plausible that D+Ea is the source of the ILR
emission, we generated a grid of models to determine if the observed
variations in the C~IV flux could result from the response of this component
to changes in the strength of the ionizing continuum. As with our model of the
emission line ratios, we assumed that the ionizing flux, and therefore $U$,
scale linearly with the continuum flux at 1350 \AA. Figure 3 shows that Low-N,
with $N_{H}$ = 10$^{22.46}$ cm$^{-2}$, cannot replicate the behavior of the
observed C~IV fluxes, as the model curve turns over too quickly, due to the
fact that too much C$^{+3}$ is ionized into C$^{+4}$ at the higher flux
states.

We were able to remedy this problem by increasing $N_{H}$ to 10$^{22.93}$
cm$^{-2}$, the column density of D+Ea in 2000 March (Table 1). In Table 3,
we list the emission line ratios for this model (``High-N''). The differences in
line ratios between the low- and high column density models are negligible.
Figure 3 shows that High-N is able to reproduce the observed positive
correlation of continuum and C IV ILR fluxes. Thus, we conclude that the higher
column density is more appropriate globally, and that the low column in 2002 May
due to transverse motion is atypical. This is consistent with our previous claim
that variations in the equivalent widths of the absorption lines from D+E over
the time period of the STIS echelle observations  were primarily due to changes
in the ionizing continuum, until the large drop in column density seen in 2002
May (Paper II). Interestingly, we found that varying the ionization parameter
for X-High suppresses $F_{CIV}$ for the lowest values of $U$, which suggests
that the ILR may not be fully screened by this component.

The global covering factor ($C_g$) of the ILR gas can be obtained from the scale
factor needed to match the model curve in Figure 3 with the general trend of the
observed C~IV fluxes. For model High-N, we find that $C_g =$ 0.4, which is
significant, and yet reassuringly does not exceed one. The predicted C~IV flux
initially increases with continuum flux, but begins to decrease after peaking at
F(1350) = 1.3 $\times 10^{-13}$ ergs cm$^{-2}$ s$^{-1}$.  Thus, our highest
observed C~IV flux comes after the peak, and the model predicts that even higher
continuum levels would result in significantly lower C~IV fluxes from the ILR.

Although the High-N model matches the behavior of the ILR C~IV emission
reasonably well, the agreement is not perfect. We attribute these discrepancies
to the size of the ILR and light travel time effects, which are difficult to
disentangle since the light curves in Figure 2 are severely undersampled. For
example, the most discrepant C~IV flux, obtained during 2002 May, may simply be
due to the fact that the higher continuum flux in our line-of-sight has yet to
reach the bulk of the ILR gas, which is irradiated by weaker continuum flux from
an earlier epoch as we see it. Another possible problem is that the global
covering factor required for the ILR models exceeds, by a factor of $\sim$ 4,
the constraint that we derived from the forbidden O~VII 22.1 \AA~ emission line
in Paper I. A likely explanation is that most of the O~VII emission is absorbed
across the entire profile by the large soft X-ray opacity (primarily the He~II
edge) of D+Ea, whereas the UV resonance lines are only self-absorbed in part of
their blue wings.

\section{Summary and Discussion}

We have discovered a distinct component of the emission lines in NGC~4151 that
has a width (FWHM $\approx$  1170 km s$^{-1}$) between those of the classic
broad and narrow emission lines. We were able to isolate this component by using
{\it HST} spectra at high spectral and angular resolutions, which reduce the
contribution of the NLR to the lines, of NGC~4151 in a low continuum-flux state,
which reduces the BLR contribution. This intermediate-line region (ILR) would be
difficult to detect in ground-based observations, especially in moderate- to
high-flux states, and its connection to previous claims of ILRs in Seyfert
galaxies and quasars (e.g., Brotherton et al. 1994; cf. Sulentic \& Marziani
1999) is therefore uncertain.

The properties of the ILR lead us to believe that we have found the emission
from at least the major component (D+Ea) of the high-column outflowing absorbers
in NGC~4151, based on the following evidence. 1) The ILR emission and
blueshifted absorption lines extend to about the same velocity range ($-$1400 km
s$^{-1}$), and together they resemble P-Cygni profiles, suggesting
self-absorption. 2) The approximate location of the ILR ($\sim$0.1pc), derived
from the width of the emission lines, is consistent with that of the D+E
absorbers based on their ionization parameters and densities. 3) A
photoionization model of the D+Ea absorber that is completely constrained in
terms of $U$, $N_H$, and $n_H$ gives line ratios that are close to the observed
ILR values.

Given that the ILR emission lines are from the same kinematic component that
produces the D+Ea absorption, the emission lines provide additional constraints
on the mass outflow. Our photoionization models are able to reproduce the
observed correlation of continuum and C~IV fluxes in the STIS data if we use the
higher column density from 2000 March ($N_H$ $=$ 10$^{22.93}$ cm$^{-2}$ ), as
opposed to that from 2002 May ($N_H$ $=$ 10$^{22.5}$ cm$^{-2}$), suggesting that
the former value is a more appropriate global average. Furthermore, the scale
factor needed to match the observed ILR C~IV fluxes gives us a global covering
factor of $C_g$ $=$ 0.4 for the bulk of the mass outflow. This value derived for
an individual Seyfert is similar to that derived from the frequency of absorbers
in Seyfert galaxies, $C_g$ $\approx$ 0.5, assuming that all Seyfert 1 galaxies
have mass outflows (as opposed to $\sim$50\% of Seyferts having outflows with
C$_g$ $=$ 1, Crenshaw al. 1999). However, we note that the D+Ea absorption in
NGC~4151 has an unusually large column density and small distance from the
continuum source compared to most UV absorbers (Crenshaw et al. 2003). One
possible interpretation is that the majority of the mass outflow in NGC~4151 is
at an early stage of evolution. Another possibility is that the high column is
due to a special viewing angle, close to the edge of the NLR bicone, as
discussed below.

Previous photoionization studies (Alexander et al. 1999; Kraemer et al. 2000)
have concluded that the ionizing radiation incident on the NLR in NGC~4151 is
filtered by absorption close to the continuum source. In Kraemer et al.
(2000), we predicted an absorber with $U = 1$ and $N_H = 10^{22.5}$, which are
remarkably close to the values for D+Ea and the ILR. Thus, the ILR is likely
shielding the NLR, which places some interesting constraints on its geometry.
Based on kinematic models of {\it HST} long-slit spectra of NGC~4151, we found
that the velocity field of the NLR can be explained by radial outflow in a
bicone with an axis inclined by $\sim$45\deg\ with respect to our line of
sight, with the outflowing gas roughly confined between inner and outer
half-opening (polar) angles of $\theta =$ 15\deg\ and 33\deg\ (Das et al.
2005). Although this geometry places our line of sight just outside of the
bicone, we can still see the nucleus of this Seyfert 1 galaxy because the
source of obscuration, and therefore the bicone edges, do not have sharp edges
(Evans et al. 1993; Schmitt et al. 2006). In order to shield the NLR and be in
our line of sight as well, the ILR must cover at least polar angles from
15\deg\ to 45\deg\, which yields a global covering factor of $C_g \geq 0.26$.
Extending the coverage to $\theta =$ 15 -- 55\deg\ increases $C_g$ to the
observed value of 0.4, whereas filling in the cone so that $\theta =$ 0 --
55\deg\ increases it slightly more to 0.43. We conclude that mass outflow in
the ILR covers a large solid angle that is likely focused in the polar
direction. Furthermore, the high column density in our line of sight ($\theta
= 45$\deg) compared to other Seyferts that are more likely viewed pole-on,
suggests that the $N_H$ is increasing with polar angle, and assuming this
trend continues, it is consistent with previous suggestions that the source of
obscuration in the unified scheme is an optically thick wind more focused
toward the plane of the accretion disk ($\theta = 90$\deg) (K\"{o}nigl \&
Kartje 1994; Elitzur \& Shlosman 2006).

In Papers I and II, we determined transverse velocities of the D+E absorbers
that were on the order of $v_T \approx 2100$ km s$^{-1}$ or larger. These
values are considerably higher than the radial velocity centroid at $v_r =
-490$ km s$^{-1}$, which suggests that the kinematics of the gas at $\sim$0.1
pc is dominated by rotation, but with a significant outflow component. A
transverse velocity of 2100 km s$^{-1}$ leads to a somewhat smaller distance
($\sim$0.04 pc), but that assumes purely gravitational motion at this
distance, which may not be the case, as discussed below.

It is interesting that the transverse velocity of the absorber is higher than
the largest emission-line radial velocity, given by the half-width at zero
intensity (HWZI) of the emission lines ($\sim$1400 km s$^{-1}$), which is
possible for a number of different outflow geometries. For example, in Figure
4, we show perhaps the simplest model that can satisfy our geometric and
kinematic constraints. In this model, the line of sight velocity is the same
as the velocity in the ``r'' direction ($v_r$), which can also be equated to
the outflow velocity. For simplicity, we assume constant $v_{r}$ and
$v_{\phi}$ (in standard spherical coordinates) and that the flow is at a
constant polar angle ($v_T = v_{\phi}$, $v_{\theta} = 0$), although in general
$\bf{v_T}$ is the vector sum of $\bf{v_{\phi}}$ and $\bf{v_{\theta}}$. The
resulting curve of motion is a spiral along a cone with a half-opening angle
of $\theta$ = 45\deg. If $v_{\phi} = 2100$ km s$^{-1}$ holds for other ILR
clouds (i.e., those not seen in absorption), we calculate that the highest
emission-line radial velocity in this model is 1550 km s$^{-1}$, which is
indeed smaller than the transverse velocity and close to the observed HWZI of
the ILR profile. For comparison, we also plot a curve in Figure 4 with
$v_{\phi} = 10,000$ km s$^{-1}$, which helps to show the conical surface that
the clouds follows. However, this case results in emission-line velocities
much higher than those seen in the ILR. Thus, at least in this simple model,
$v_T$ cannot be much higher than $\sim$2100 km s$^{-1}$.

It is easy to think of D+Ea as a cloud, as drawn in Figure 4, since the radial
thickness of D+Ea is $N_H/n_H = 1.3 \times 10^{16}$ cm $= 0.004$ pc, which is
only $\sim$4\%\ of its radial distance from the central source. Thus, the ILR
can then be thought of as a collection of ``clouds'' that altogether cover
$\sim$ 40\% of the sky as seen from the central source. However, it could also
take the form of a relatively thin (partial) shell or flow tube with column
density inhomogeneities in the radial direction. The relative thickness of
X-high is $\sim$30\% if it is near D+E (but a smaller percentage at smaller
distances), so it could be more like a semi-continuous ``wind''.

Assuming that the radial velocity centroid of D+Ea is the approximate outflow
velocity, the mass outflow rate due to this component is $\dot{M}_{out} = 4 \pi
r N_H m_H C_g v_r = 0.16$ M$_{\odot}$ yr$^{-1}$, where $m_p$ is the mass of a
proton. If we include the other X-ray and UV components (dominated by X-high)
and assume the same $C_g$, the outflow rate increases to $\dot{M}_{out} = 0.19$
M$_{\odot}$ yr$^{-1}$. On the other hand, the inferred accretion rate is only
$\dot{M}_{acc} = 0.013$ M$_{\odot}$ yr$^{-1}$, based on the average bolometric
luminosity of NGC~4151 ($L_{bol} = 7.3 \times 10^{43}$ ergs s$^{-1}$, Kaspi et
al. 2005) and an expected efficiency of $\sim$0.1 in the conversion of mass
infall to energy (Peterson 1997). This difference can be explained by one of two
scenarios, or a combination of the two). 1) In a steady-state situation, most
(more than 90\%)of the inflowing mass at $\sim$0.1 pc is not eventually
accreted, but is instead driven outward, or 2) the current episode of mass
ejection in NGC~4151 is an unusual or infrequent event, preceded by a long
interval of mass accumulation into a reservoir (e.g., accretion disk).

In Paper I, we examined the possible dynamical forces (radiation driving,
thermal expansion, magnetocentrifugal acceleration) that could possibly drive
the outflow in NGC~4151. Here we review our arguments using a revised value for
the black-hole mass ($M = 4.1 \times 10^{7}$ M$_{\odot}$, Metzroth et al. 2006),
which yields an Eddington ratio of $L_{bol}/L_{E} = 0.014$ . In order for
radiation pressure to drive the gas outward from the nucleus, the force
multiplier must be $FM \geq (L_{bol}/L_{E})^{-1} = 71$. From our photoionization
models, we found that $FM \leq 2 $ for X-High, $FM \leq 41 $ for D+Ea in
2002, and $FM \leq 11 $ for D+Ea at historically high continuum flux levels
(four times the 2002 level, see Paper I). These values are upper limits because
they were calculated for the optically thin case, whereas the force multiplier
decreases into a cloud as it becomes optically thick (Chelouche \& Netzer 2005).
Dust grains are efficient absorbers of radiation and could greatly increase the
effective force multipliers, but the lack of reddening indicates that dust is
essentially absent in the high-column absorbers (e.g., a Galactic dust/gas ratio
would result in a reddening of E$_{B-V} \approx 16$ mag for D+Ea). The lack of
dust is not surprising since the absorbers are near and likely originated inside
the dust sublimation radius, $r_{sub} \approx 1.3L_{46}^{1/2}~T_{1500}^{-2.8}$
pc $\approx 0.11$ pc (Barvainis 1987). Thus, X-high cannot be radiatively driven
and D+Ea is only marginally susceptible to radiation driving in low to medium
flux states. However, we cannot rule out the possibility that the absorbers
had lower $U$ when they were closer to the central nucleus and were therefore
more susceptible to radiation driving.

For a thermal wind, the radial distance at which the gas can escape is given by
$r_{esc} \geq G M m_H / T_g k$ (Crenshaw et al. 2003, and references therein),
where $T_g$ is the gas temperature and k is the Boltzmann constant. Our
photoionization models give $T_g =$ $5 \times 10^4$ K and 3 $\times 10^{6}$ K
for D+Ea and X-High, which yield $r_{esc} \geq 1.3 \times 10^{21}$ cm and $\geq
2.2 \times 10^{19}$, respectively. These are much greater than the radial
locations of D+Ea and X-high, so these high-column components cannot be
attributed to thermal winds.

An MHD flow is therefore a likely possibility, at least by comparison to the
current alternatives. Accretion disk winds that invoke magnetocentrifugal
acceleration (Emmering, Blandford, \& Shlosman 1992; Bottorff, Korista, \&
Shlosman 2000) predict that the gas will flow along rotating magnetic field
lines, which would result in significant transverse velocities as seen by an
outside observer, as illustrated in Figure 4. In the presence of
non-dissipative MHD waves, the line widths can greatly exceed the thermal widths
(Bottorff et al. 2000), which could explain the large width of D+Ea (FWHM $=$
450 km s$^{-1}$). Everett (2005) presents an accretion-disk wind model that
incorporates both radiative and magnetocentrifugal acceleration. Further clues
to the nature of mass outflows in AGN will come from comparing predictions from
these models with detailed observational constraints, such as those we have
presented for NGC~4151.

\acknowledgments

We thank Brad Peterson, Gary Ferland, Alvin Das, and John Everett
for helpful discussions. Some of the data presented in this paper were obtained
from the Multimission Archive at the Space Telescope Science Institute (MAST).
STScI is operated by the Association of Universities for Research in Astronomy,
Inc., under NASA contract NAS5-26555. Support for MAST for non-HST data is
provided by the NASA Office of Space Science via grant NAG5-7584 and by other
grants and contracts. This research has made use of the NASA/IPAC Extragalactic
Database (NED) which is operated by the Jet Propulsion Laboratory, California
Institute of Technology, under contract with the National Aeronautics and Space
Administration.

\clearpage

\figcaption[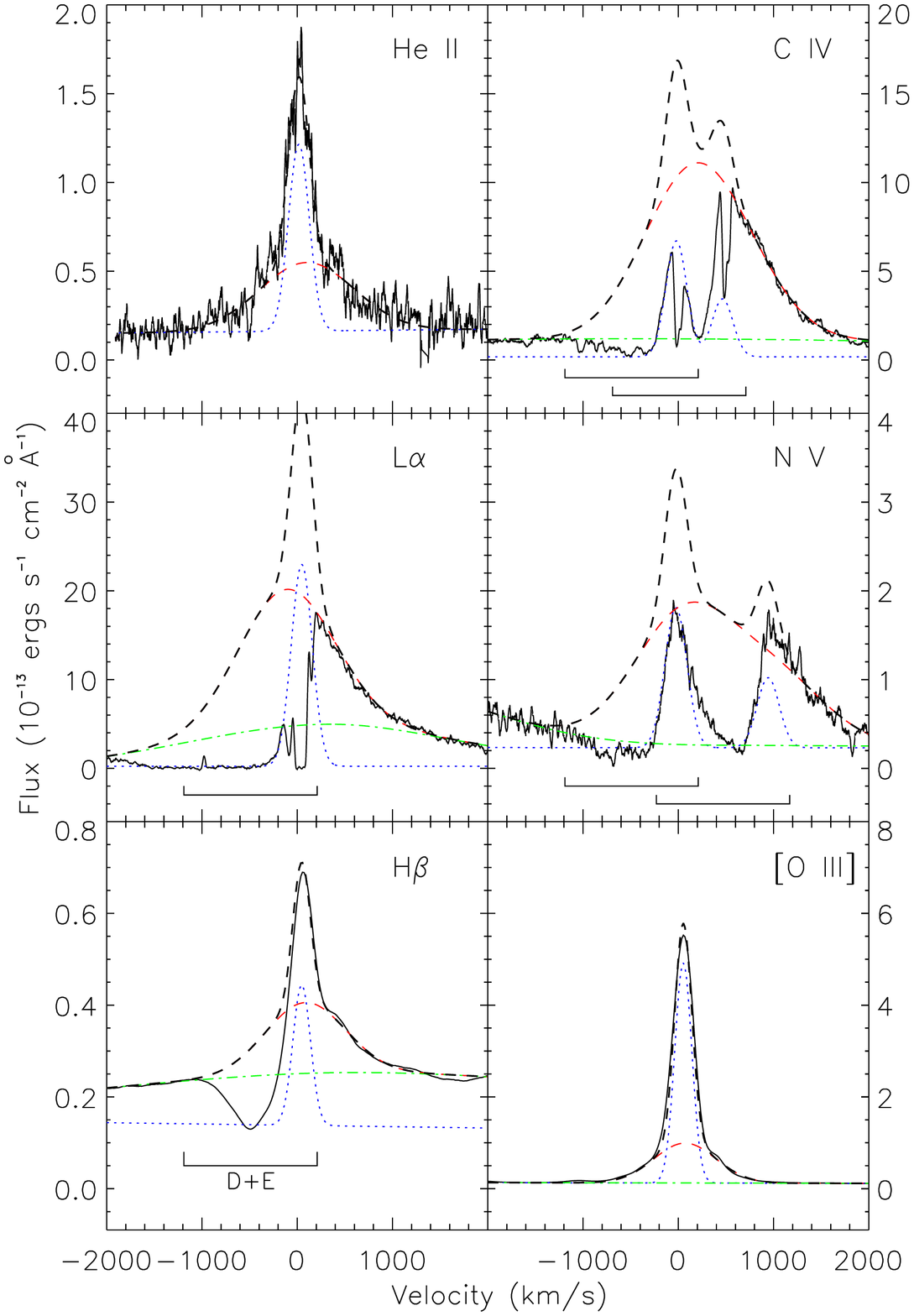]{Emission-line profiles from the STIS E140M observation on
2000 June 15 (top four panels) and G430M observation (bottom two panels) on 2000
July 2. The solid black line shows the observed flux as a function of radial
velocity in the rest frame of NGC~4151 (z $=$ 0.00332).  Components of the
profiles fits are: continuum + NLR (dotted blue), continuum + BLR (dotted-dashed
green), continuum + BLR + ILR (dashed red), and continuum + BLR + ILR + NLR
(dashed black). The full extent of absorption component D+E (including doublets)
is given below the profiles. The absorption at more negative velocities in the
L$\alpha$ profile is due to Galactic L$\alpha$, and the absorption in the cores
of the narrow UV resonance lines is due to components F and F$'$ from NGC 4151's
ISM and halo. The slight offsets in the apparent velocities of the narrow and
intermediate profiles are due to blending of the doublets and/or to
different shapes and contributions of the underlying broad and continuum
emissions.}

\figcaption[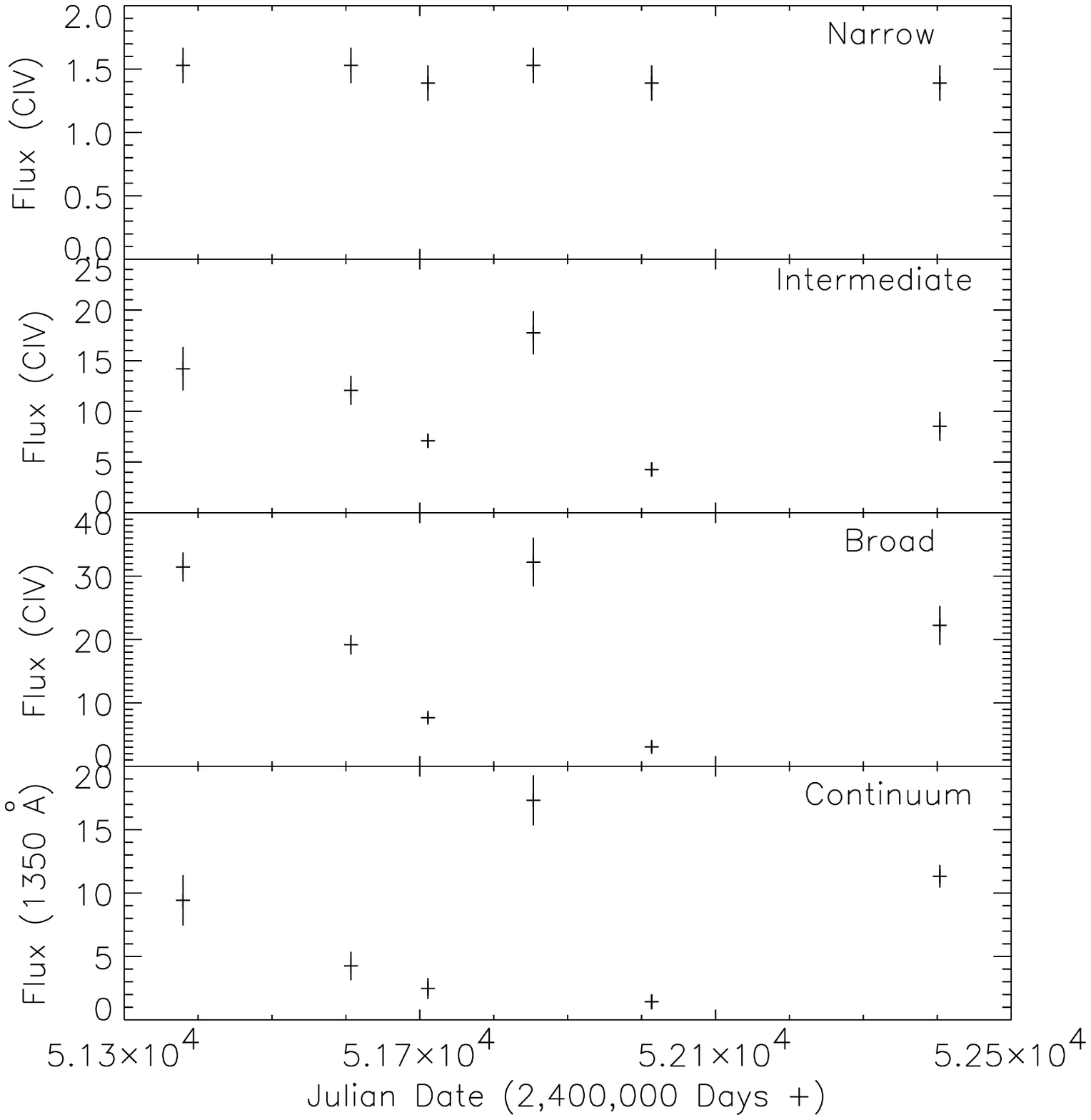]{Light curves of the continuum flux at 1350 \AA\ (bottom
panel, in units of 10$^{-14}$ ergs s$^{-1}$ cm$^{-2}$ \AA$^{-1}$) and the C~IV
emission components (top three panels, in units of 10$^{-12}$ ergs s$^{-1}$
cm$^{-2}$) from the STIS E140M spectra.}

\figcaption[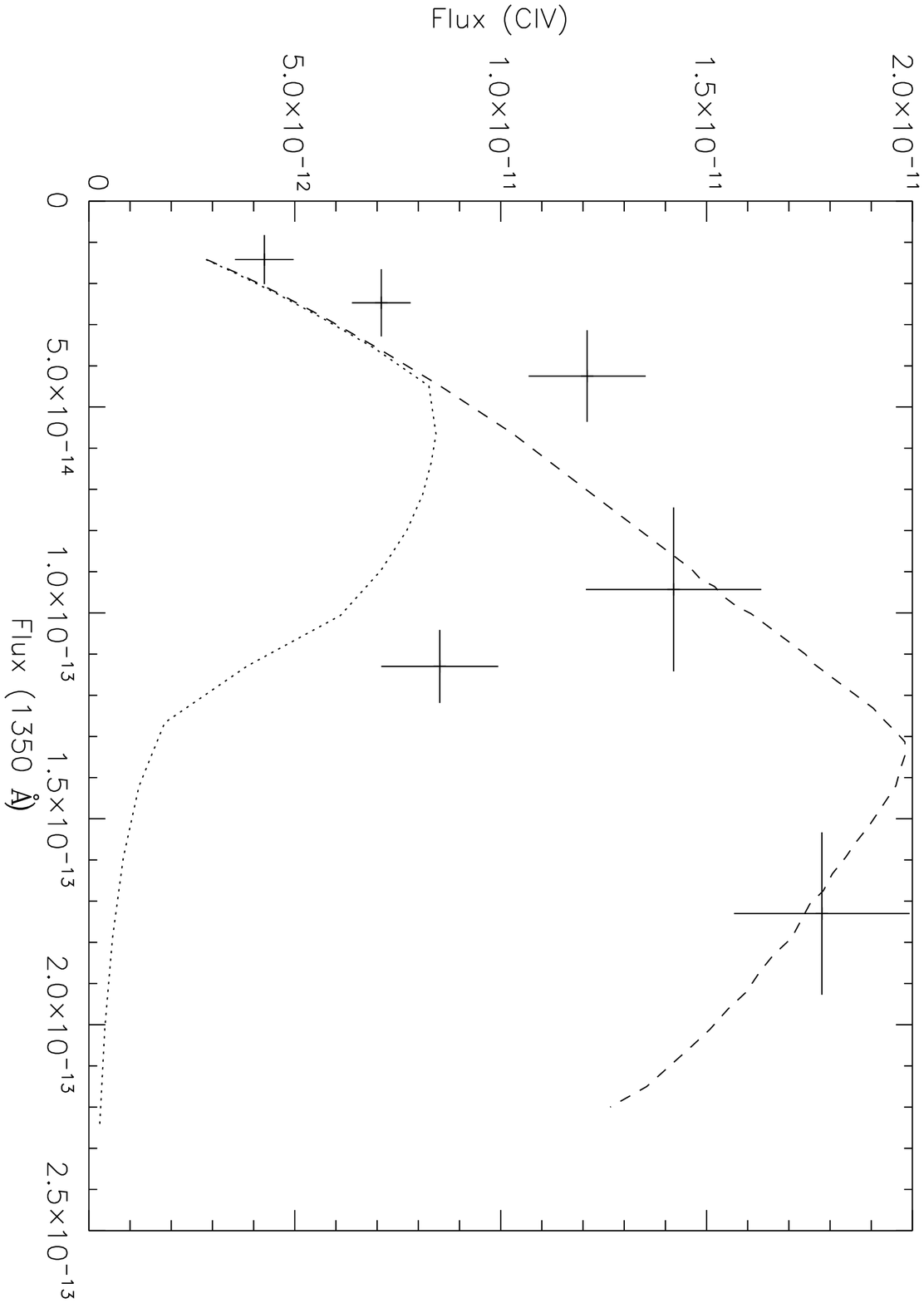]{The model predictions of ILR C~IV flux as a function of
continuum flux at 1350 \AA\, for $N_{H} = 10^{22.46}$ cm$^{-2}$ (dotted line)
and $N_{H} = 10^{22.93}$ cm$^{-2}$ (dashed line), compared to the observed
fluxes. The details of the scaling of the model results are given in the text.
The continuum flux is in units of ergs s$^{-1}$ cm$^{-2}$ \AA$^{-1}$ and the
C~IV emission flux is in units of ergs s$^{-1}$ cm$^{-2}$).}

\figcaption[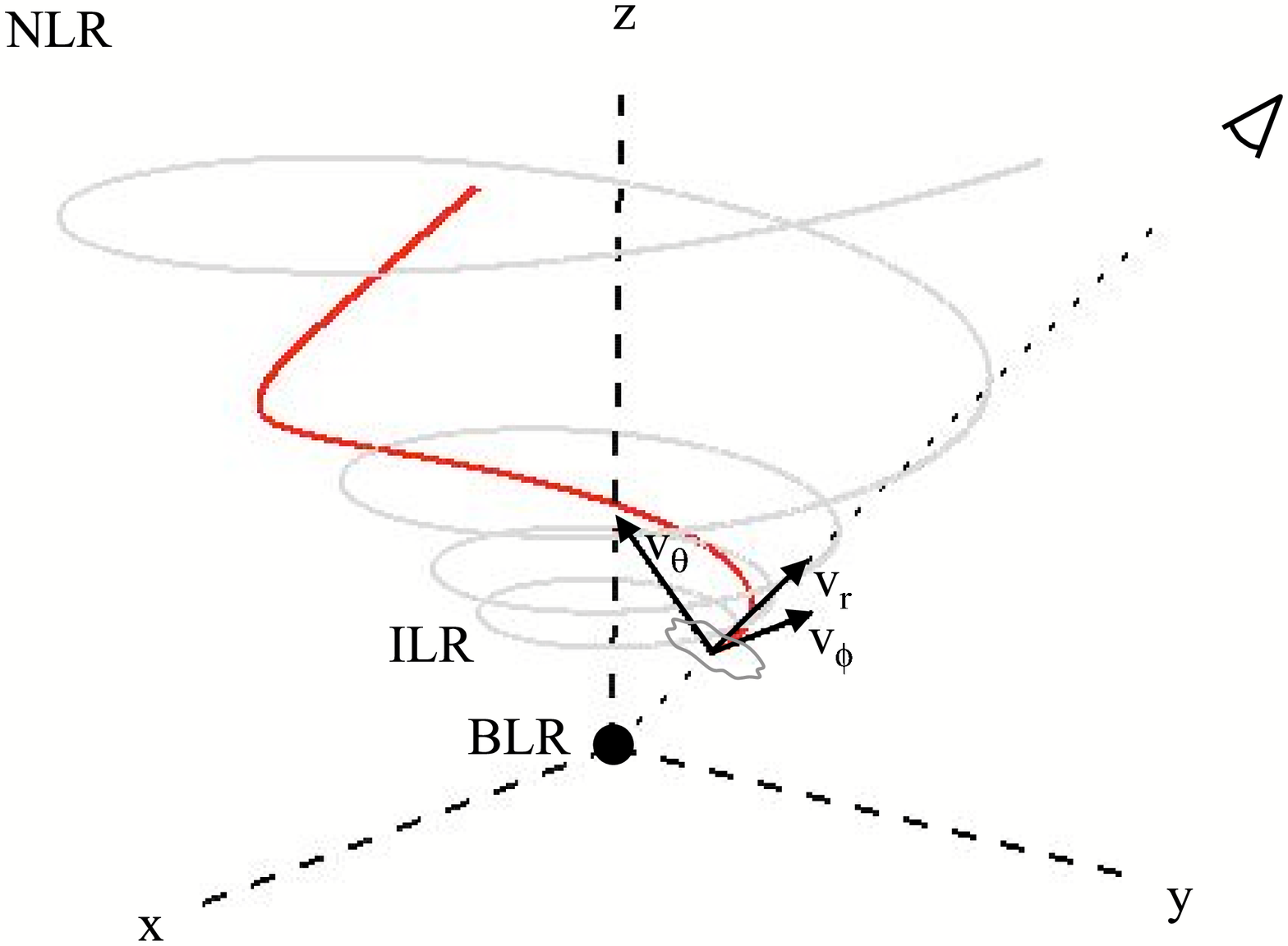]{Schematic diagram of a simple model for the geometry and
kinematics of the ILR. The line of sight is in the y-z plane, at a polar angle
of $\theta = 45$\deg\ from the rotation axis (z) of the accretion disk. Unit
vectors for the velocity components are shown at the location of the D+Ea
absorber, at a radial distance of 0.1 pc from the inner accretion disk and BLR.
The solid red line shows the curve of motion for the D+Ea absorber over 1000
years assuming $v_{r} = 490$ km s$^{-1}$, $v_{\phi} = 2100$ km s$^{-1}$, and
$v_{\theta} = 0$. The grey curve shows the curve of motion over the same time
period for $v_{r} = 490$ km s$^{-1}$, $v_{\phi} = 10,000$ km s$^{-1}$, and
$v_{\theta} = 0$. Additional ILR clouds moving in a similar fashion are assumed
to cover the BLR at a distance of $\sim$0.1 pc within the cone at $\theta \leq
45$\deg, and possible extending to $\theta \approx 55$\deg.}

\clearpage

\begin{deluxetable}{lccccc}
\tablecolumns{6}
\footnotesize
\tablecaption{NGC 4151 Absorption Components}
\tablewidth{0pt}
\tablehead{
\colhead{Name}  & \colhead{Date} & \colhead{$C_{los}$}  & \colhead{log $U$}  &
\colhead{log $N_H$} & \colhead{log $n_H^{a}$}\\
\colhead{} & \colhead{} & \colhead{} & \colhead{} & \colhead{(cm$^{-2}$)} &
\colhead{(cm$^{-3}$)}
}
\startdata
X-high & 2002 May & 1.0 &  1.05 & 22.5  & $\geq$5.5 \\
D+Ea   & 2002 May & 0.9 & $-$0.27 & 22.46 & 6.8 \\
       & 2000 Mar & 0.9 & $-$0.57 & 22.93 & 6.8 \\
D+Eb   & 2002 May & 0.8 & $-$1.67 & 20.8  & 8.1 \\
D+Ec   & 2002 May & 0.5 & $-$1.08 & 21.6  & 7.4 \\
D+Ed   & 2002 May & 0.2 & $-$3.1 & 19.5  & 9.5 \\
       & 2001 Apr & 0.7 & $-$4.0 & 19.5  & 9.5 \\
\enddata
\tablenotetext{a}{From profile fits to the metastable C~III absorption
for subcomponents a, b, and c (Paper II). The number density for D+Ed is
obtained by assuming co-location with the other subcomponents, and the upper
limit for X-high is derived from evidence that it is interior to D+E.}
\end{deluxetable}

\begin{deluxetable}{lcc}
\tablecolumns{3}
\footnotesize
\tablecaption{STIS E140M Spectra of NGC 4151}
\tablewidth{0pt}
\tablehead{
\colhead{UT Date} & \colhead{Julian Date} & \colhead{F(1350)$^a$}
}
\startdata
1999 July 19     & 2,451,380  &  9.4 \\
2000 March 3     & 2,451,607  &  4.3 \\
2000 June 15     & 2,451,711  &  2.5 \\
2000 November 5  & 2,451,854  & 17.3 \\
2001 April 14    & 2,452,014  &  1.4 \\
2002 May 8       & 2,452,403  & 11.3 \\
\enddata
\tablenotetext{a}{Units of 10$^{-14}$ ergs s$^{-1}$ cm$^{-2}$ \AA$^{-1}$.}
\end{deluxetable}

\begin{deluxetable}{lrrcc}
\tablecolumns{5}
\footnotesize
\tablecaption{NGC 4151 -- ILR Line Ratios
(Relative to H$\beta$$^a$) 
\label{tbl-1}}
\tablewidth{0pt}
\tablehead{
\colhead{Emission Line} & \colhead{Observed}& \colhead{Dereddened$^b$}
& \colhead{Model Low-N$^c$} & \colhead{Model High-N$^d$}
}
\startdata
Ly$\alpha$ $\lambda$1216  & 41.7 & 46.9$\pm$13.2 & 42.3 & 39.4  \\
N~V $\lambda$1240         &  6.2 &  6.9$\pm$1.9  & 2.9 & 2.7\\
C~IV$\lambda$1550         & 37.5 & 40.6$\pm$9.0 & 28.7 & 26.7 \\
He~II $\lambda$1640        &  1.4 &  1.5$\pm$0.4 & 1.9 & 1.8\\
$[$O~III] $\lambda$5007   &  4.5 &  4.5$\pm$0.8 & 5.4 & 5.1
\enddata
\tablenotetext{a}{Observed flux of ILR H$\beta$ $=$ 1.89 ($\pm$0.32) $\times$
10$^{-13}$ ergs s$^{-1}$ cm$^{-2}$}
\tablenotetext{b}{From E$_{B-V}$ $=$ 0.02 and the standard Galactic reddening
curve of Savage \& Mathis (1979).}
\tablenotetext{c}{Model parameters: $U = 10^{-0.92}, N_{H} = 10^{22.46}$
cm$^{-2}$, and  $n_{H} =
10^{6.75}$ cm$^{-3}$.}
\tablenotetext{d}{Model parameters: $U = 10^{-0.92}, N_{H} = 10^{22.93}$
cm$^{-2}$, and  $n_{H} =
10^{6.75}$ cm$^{-3}$.}
\end{deluxetable}


\clearpage
\begin{figure}
\plotone{f1.eps}
\\Fig.~1
\end{figure}

\clearpage
\begin{figure}
\plotone{f2.eps}
\epsscale{1.0}
\\Fig.~2
\end{figure}

\clearpage
\begin{figure}
\plotone{f3.eps}
\\Fig.~3
\end{figure}

\clearpage
\begin{figure}
\plotone{f4.eps}
\\Fig.~4
\end{figure}


\clearpage
\begin{references}

\reference{ale1999}Alexander, T., Sturm, E., Lutz, D., Sternberg, A., Netzer,
H., \& Genzel, R. 1999, \apj, 512, 204.

\reference{bar1987}Barvainis, R. 1987, \apj, 320, 537.

\reference{bro1985}Bromage, G.E., et al. 1985, \mnras, 215, 1.

\reference{bot2000}Bottorff, M.C., Korista, K.T., Shlosman, I. 2000, \apj, 537,
134.

\reference{bro1994} Brotherton, M.S., Wills, B.J., Francis, P.J., \& Steidel,
C.C. 1994, \apj, 430, 495.

\reference{che2005}Chelouche, D. \& Netzer, H. 2005, \apj, 625, 95.

\reference{cre2005}Crenshaw, D.M. \& Kraemer, S.B. 2005, \apj, 625, 680.

\reference{cre1999}Crenshaw, D.M., Kraemer, S.B., Boggess, A., Maran, S.P.,
Mushotzky, R.F., \& Wu, C.-C. 1999, \apj, 516, 750.

\reference{cre2003}Crenshaw, D.M., Kraemer, S.B., \& George, I.M. 2003, \araa,
41, 117.

\reference{cre2001}Crenshaw, D.M., et al. 2000, \apj, 545, L27.

\reference{das2005}Das, V., et al. 2005, \aj, 130, 945.

\reference{emm1992}Emmering, R.T., Blandford, R.D., \& Shlosman, I. 1992,
\apj,
385, 460.

\reference{eli2006} Elitzur, M. \& Shlosman, I. 2006, \apj, in press
(astro-ph/0605686).

\reference{eva1993}Evans, I.N., Tsvetanov, Z., Kriss, G.A., Ford, H.C.,
Caganoff, S., \& Koratkar, A.P. 1993, \apj, 417, 82.

\reference{eve2005}Everett, J.E. 2005, \apj, 631, 689.

\reference{fer1998}Ferland, G.J., Korista, K.T., Verner, D.A., Ferguson, J.W., 
Kingdon, J.B., \& Verner, E.M. 1998, \pasp, 110, 749.


\reference{ham2002}Hamann, F., Korista, K.T., Ferland, G.J., Warner, C., \&
Baldwin, J. 2002, \apj, 564, 592.

\reference{hut2002}Hutchings, J.B., Crenshaw, D.M., Kraemer, S.B., Gabel, J.R.,
Kaiser, M.E., Weistrop, D., \& Gull, T.R. 2002, \aj, 124, 2543.

\reference{kas2005}Kaspi, S., Maoz, D., Netzer, H., Peterson, B.M.,
Vestergaard, M., \& Jannuzi, B.T. 2005, \apj, 629, 61.

\reference{kon1994}K\"{o}nigl A., \& Kartje J. 1994, \apj, 424, 446.

\reference{kra2000b}Kraemer, S.B. \& Crenshaw, D.M. 2000, \apj, 544, 763.

\reference{kra2000b}Kraemer, S.B., Crenshaw, D.M., Hutchings, J.B., Gull, T.R.,
Kaiser, M.E., Nelson, C.H., \& Weistrop, D. 2000, \apj, 531, 278.

\reference{kra2004}Kraemer, S.B., Ferland, G.J, \& Gable, J.R. 2004, \apj, 604,
56.

\reference{kra2001}Kraemer, S.B., et al. 2001, \apj, 551, 671.

\reference{kra2005} Kraemer, S.B., et al. 2005, \apj, 633, 693 (Paper I).

\reference{kra2005} Kraemer, S.B., et al. 2006, \apj, in press
(astro-ph/0608383) (Paper II).

\reference{kro2001}Krolik, J.H. \& Kriss, G.A. 2001, \apj, 561, 684.

\reference{met2006}Metzroth, K.G., Onken, C.A., \& Peterson, B.M. 2006, \apj,
647, 901.

\reference{pet1993}Peterson, B.M. 1993, \pasp, 105, 247.

\reference{pet1997}Peterson, B.M. 1997, An Introduction to Active Galactic
Nuclei (Cambridge, UK: Cambridge University Press).

\reference{pet2004}Peterson, B.M., et al. 2004, \apj, 613, 682.

\reference{pro2003}Proga, D. 2003, \apj, 585, 406.

\reference{sch2006}Schmitt, H.R., Kraemer, S.B., Crenshaw, D.M., \& Hutchings,
J.B. 2006, in The X-ray Universe 2005, ed. A. Wilson, ESA SP-604, 661.

\reference{sul1999}Sulentic, J.W. \& Marziani, P. 1999, \apj, 518, 9.

\reference{wey1997}Weymann, R.J., Morris, S.K., Gray, M.E., \& Hutchings, J.B.
1997, \apj, 483, 717.

\end{references}
\end{document}